\documentclass[aip,twocolumn,reprint,showpacs,superscriptaddress]{revtex4-1}
\usepackage{}
\usepackage{txfonts}
\usepackage{amssymb}
\usepackage{graphicx}
\usepackage{color}
\begin{document}

\title{Two stage $\gamma$ ray emission via an ultrahigh intensity laser pulse interaction with a laser-wakefield accelerated electron beam}
\author{M. A. Bake}
\email[Author to whom correspondence should be addressed. Electronic mail:]{mabake@xju.edu.cn.}
\affiliation{School of Physics Science and Technology, Xinjiang University, Urumqi 830046, People's Republic of China}
\affiliation{Center of Theocratical Physics, Xinjiang University, Urumqi 830046, People's Republic of China}
\author{A. Tursun}
\affiliation{School of Physics Science and Technology, Xinjiang University, Urumqi 830046, People's Republic of China}
\author{A. Aimidula}
\affiliation{School of Physics Science and Technology, Xinjiang University, Urumqi 830046, People's Republic of China}
\author{B. S. Xie}
\affiliation{College of Nuclear Science and Technology, Beijing Normal University, Beijing 100875, People's Republic of China}
%\date{\today}

\begin{abstract}
We investigate the generation of twin $\gamma$ ray beams in collision of an ultrahigh intensity laser pulse with a laser wakefield accelerated electron beam by using particle-in-cell simulation. We consider the composed target of a homogeneous underdense preplasma in front of an ultrathin solid foil. The electrons in the preplasma are trapped and accelerated by the wakefield. When the laser pulse is reflected by the thin solid foil, the wakefield accelerated electrons continue to move forward and passing through the foil almost without the influence of the reflected laser pulse and the foil. Consequently, two groups of $\gamma$ ray flashes, with tunable time delay and energy, are generated by the wakefield accelerated electron beam interacting with the reflected laser pulse from the foil as well as another counter propagating petawatt laser pulse in the behind the foil. The dependence of the $\gamma$ photon emission on the preplasma densities, driving laser polarization and the foil are studied.
\end{abstract}
\pacs{52.38.Kd, 52.38.Hb, 52.65.Rr}
\maketitle
\section{Introduction}
The idea of laser-wakefield accelerators (LWFAs) are first proposed by the Tajima and Dawson four decades ago \cite{Tajima}. At that time, they promise to provide high-energy compact electron sources due to its tremendous accelerating gradients, which cannot be delivered by the conventional radio frequency (RF) accelerators. In RF the metal cavity have electric break down limit; they cannot support electric field gradients greater than 100 MV/m. However, the plasma have not the electric break down limit and can support more than 100 MV/m gradients \cite{Malka}. Recent related researches indicate that the electron beams with energies of several GeV could be obtained by LWFAs \cite{Wang,Leemans,Gonsalves}. Such an energetic electrons will be provided novel compact light sources under the ultrahigh intensity laser conditions \cite{Powers,Lemos,Chenmin}, which have some applications in the  medicine, biology, industry, condensed matter, high energy density science and so on \cite{Albert}.

With the rapid advancement of the ultrahigh intensity laser technologies and related experimental conditions, some new developed Petawatt (PW) laser facilities could deliver ultrashort laser pulses with the intensities up to $10^{23}-10^{24}\mathrm{W/cm^2}$ \cite{Petawatt,ELI,ican,vulcan}. One of the most important applications of such an intense laser pulses are the generation of bright $\gamma$ rays thorough QED processes in the laser plasma or laser electron beam interactions\cite{Benedetti,Yan,Vranic,BakeQED}. Because of their potential applications in many fundamental researches, the considerable efforts has been made to investigate the laser-driven compact $\gamma$ ray sources and electron-positron pair productions. \cite{Chang,Yu,Sarri,Chen,Lobet,Xie}. Therefore, various schemes to get compact, tunable, flexible and bright $\gamma$ ray sources are under investigation and explain the underlying physical processes. The most extensively studied laser-based $\gamma$ ray source schemes are the laser-driven bremsstrahlung emission \cite{Cipiccia1}, synchrotron-like radiation \cite{Cipiccia2}, and all-optical $\gamma$ ray source \cite{Phuoc,Wenz}.

In order to reach the goal of making compact and tunable $\gamma$ ray source, some researchers continued their studies and many kinds of methods have been suggested on synchrotron-like radiation and all-optical $\gamma$ ray source. Because these schemes are still highly promising candidates to compact $\gamma$ ray sources based on currently available laser systems. Among them Ta Phuoc \textit{et al}. first studied the all-optical Compton $\gamma$ ray source \cite{Phuoc}. They predicted the possibility of high energy $\gamma$ rays through the simple Compton backscattering in a laser-plasma accelerator. Chen \textit{et al}. experimentally investigated the method of $\gamma$ ray generation from inverse Compton scattering laser wakefield accelerated electrons \cite{Chen}. Recently, Yan \textit{et al}. studied the nonlinear regime of Thomson scattering in the laboratory by using intense laser interacting with laser wakefield accelerated electrons \cite{Yan}. Lobet \textit{et al}. studied the pair production process in the collision of a laser-accelerated electron with a multipetawatt lasers \cite{Lobet}. Liu \textit{et al}. discussed an efficient method to generate $\gamma$ rays/hard X-rays by using double-layer target via all-optical process \cite{Liupop}. Recently, Gu \textit{et al}. reported some fruitful research results about $\gamma$ ray emission and pair production \cite{Gu1,Gu2} by laser plasma interactions. One of the our recent study also indicated that electron beams with energy of 500MeV to 1GeV, which can be obtained by current LWFAs, interacting with ultrahigh intensity laser can produce high energy photons and positrons \cite{BakeQED}. All of these studies mentioned above as well as many other researches \cite{Thomas,JMCole,Poder,Liu} are theoretically, numerically and experimentally predicted that LWFAs might be abundant sources of the $\gamma$ rays and positrons in the laboratory.

In this paper, we numerically investigate the $\gamma$ ray emission in the strong laser interaction with gas-foil compound target by using PIC simulations. In our method, energetic electrons can be produced in bubble during an intense laser propagating in the underdense preplasma. Then the solid foil, which placed behind the preplasma, reflect the driving laser pulse and interacting with the laser wakefield accelerated electron beam (first stage). As the time goes, the electron beam can passing through the foil and collide with another counter propagating high intensity laser pulse (second stage). As a results, two $\gamma$ ray beams are produced at the front and back side of the solid foil with different length and a time delay. We will study the influences of the laser polarization, preplasma densities and the foil on the $\gamma$ ray emission efficiency.

The paper is organized as follows. Section \ref{Sec2} outlines the basic target configurations and simulation parameters. The simulation results of the wakefield electron acceleration process are shown in Section \ref{Sec3}. In the beginning of the Section \ref{Sec4} the mechanism of $\gamma$ ray emission are discussed briefly, and two stage $\gamma$ ray emission process by wakefield electron interacting with a counter-propagating laser pulse are examined in detail. Among them, the driving laser pulse polarization, preplasma density and the solid foil on the $\gamma$ ray emission efficiency are also taken into account. Lastly, a brief discussion and summary are given in Section \ref{Sec5}.

\section{pic simulation model and stup}\label{Sec2}
\begin{figure}[tbp]\suppressfloats
\centerline{\includegraphics[width=9cm]{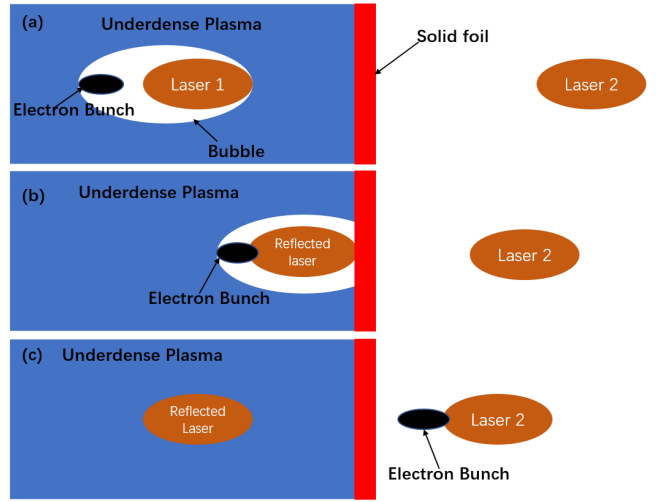}}
\caption{\label{fig1} The schematic diagram of our proposed scheme to the $\gamma$ ray production. (a) An intense laser interacting with underdense preplasma and energetic electrons are generated in bubble. (b) The laser is reflected by the solid foil and interacted with the forward LWFA electron beam in the bubble. (c) The electron beam pass inertially through the foil and collide with another counter propagating high intensity laser pulse. }
\end{figure}

In order to demonstrate our scheme, we carried out sets of two-dimensional (2D) QED-PIC simulations using the code EPOCH \cite{EPOCH}. In simulations, we consider a composed target of a homogeneous low-density preplasma in front of an ultrathin solid foil, as showed in Fig. \ref{fig1}. The preplasma is located from $x=10 \mathrm{\mu m}$ to $x=70 \mathrm{\mu m}$ in $x$ direction and from $y=-25 \mathrm{\mu m}$ to $y=25 \mathrm{\mu m}$ in $y$ direction. The thickness of the solid foil is $1 \mathrm{\mu m}$. We consider different kinds of preplasma densities while the foil density fixed to $n=30n_c$, where $n$ is the electron density and $n_c=m\omega^2/4\pi e^2$ is the critical density corresponding to the incident laser pulse, $\omega$ is the laser frequency, $m$ is the electron mass, $e$ is the electron charge.

For the LWFA the linear/circular polarized Gaussian laser pulse (driving laser pulse) are used with wavelength $\lambda=1 \mathrm{\mu m}$, spot size $y_0=6 \mathrm{\mu m}$, and the normalized amplitude is $a_0=20$. Another counter propagating laser pulse (colliding laser pulse) has a normalized amplitude of $a_0=500$ while other parameters are same as driving laser pulse. In simulation, two laser pulses enter the simulation box from the left and right boundaries with a time delay, and propagate along the positive and negative $x$ direction respectively.

Some other simulation parameters are chosen as the following: the width and height of the simulation box are $120 \mathrm{\mu m}$ and $50 \mathrm{\mu m}$ respectively, which corresponding $3600\times1000$ grid cells. 20 macro particles for preplasma and 50 macro particles for solid foil are setting in each grid cell. The periodic and simple-outflow boundary conditions in transverse and longitudinal directions are used in the simulations.

\section{Laser wakefield electron acceleration}\label{Sec3}

\begin{figure*}[tbp]\suppressfloats
\centerline{\includegraphics[width=10cm]{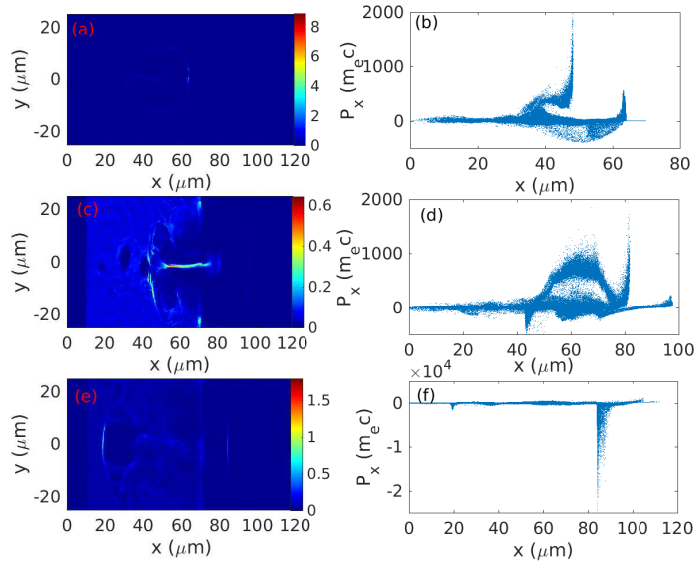}}
\caption{\label{fig2} Spatial distribution of the electron densities in the ($x, y$) plane at $t=225$ fs (a), $t=337$ fs (c), $t=420$ fs (e), respectively; and electrons phase space portrait ($x, p_x$) $t=225$ fs (b), $t=337$ fs (d), $t=420$ fs (f), respectively.}
\end{figure*}
We will investigate the $\gamma$ ray emission by a wakefield accelerated electron beam interacting with a reflected driving laser pulse and another strong laser pulse behind the foil in this paper. Moreover, it well known that the LWFA is the best way to get a high-energy electron beams in laser plasma interaction. Therefore, in a first stage, we show our simulation results of the wakefield electron acceleration.

Figure \ref{fig2} (a) shows the distribution of the preplasma electron density at $t=225$ fs for $n_e=0.05n_c$. The results indicate that the preplasma electrons are pondermotively pushed away from the laser focal areas and plasma bubble is developed. Consequently, some electrons are trapped and accelerated. From the phase space distribution of the electrons, in Fig. \ref{fig2} (b), we can observe two groups of accelerated electrons. One group by pondromotive accelerated electrons in front of the laser pulse (right pick) while another one is the wakefield accelerated electrons (left pick). These results show that an electron beam with electron density about $n_e=3n_c$ is created by the LWFA process.

The Fig. \ref{fig2} (c) and \ref{fig2} (d) show the distribution of the preplasma electrons and corresponding phase space portrait at $t=337$ fs. The wakefield accelerated electrons interacting with the reflected laser field from the solid foil, the electrons is still traveling along the positive $x$ direction, passing through the solid foil and enter the back vacuum area with beam radius of about $r_e=2 \mathrm{\mu m}$ and density of $n_e=0.4n_c$. We can see that, as the laser pulse reflected from the foil, bubble structure is destroyed, this results the electron beam escape from deceleration phase easily. Consequently, the wakefield accelerated electron beam going through the foil with narrow emittance and almost the same momentum as in the bubble, as shown in Fig \ref{fig2} (c) and \ref{fig2} (d). It is found that the wakefeild accelerated electrons are less affected neither by the reflected laser pulse nor the solid foil, and they appear at the back of the foil as a narrow energetic electron beam. This electron beam are used to $\gamma$ ray generation colliding with another counter propagating laser pulse in our present scheme, and the details shall discussed in next section.

Figure \ref{fig2} (e) and (f) are presented the simulation results of another counter propagating high-intensity laser pulse interacting with the electron beam behind the foil at $t=420$ fs. One can be found that very thin electron sheet, with density of $n_c=0.5n_c$ and height about $9\mu$m, are appeared. The phase-space portrait of longitudinal component of the electrons momentum are plotted in Fig. \ref{fig2} (f). It shows that the electrons momentum is negative since the electrons decelerated and stopped by the counter-propagating laser pulse in the second stage, change its direction to negative $x$ direction and accelerated by colliding laser pulse again. Note that, in our simulations the colliding time is choose as the electron beam colliding with laser pulse before it expands by Coulomb repulsion in order to enhance the efficiency of  $\gamma$ ray emission.
\begin{figure*}[tbp]\suppressfloats
\centerline{\includegraphics[width=10cm]{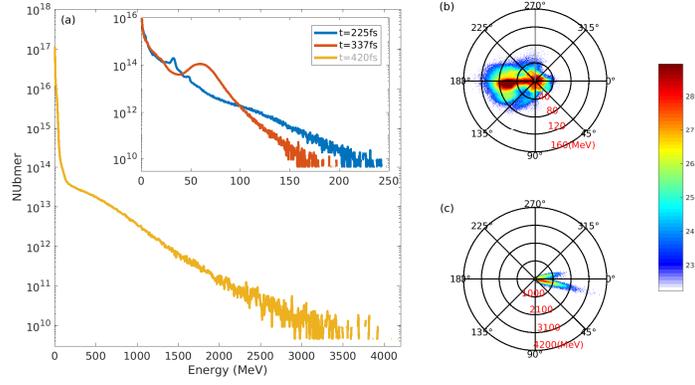}}
\caption{\label{fig3} The energy spectrum of the electrons at $t=225$ fs (blue curve), $t=337$ fs (red curve), and $t=420$ fs (yellow curve), respectively, are shown in frame \textbf{a}. The angular energy distributions of electron beam at t=225 fs and t= 420 fs are shown in frames \textbf{b} and \textbf{c}, respectively. The radial direction represents the energy and the color bar shows the electron number in $\log$ scale. }
\end{figure*}

Figure \ref{fig3}(a) is plotted the energy spectra of electron beam at different times. We can see from the Fig. \ref{fig3} (a) that an electron beam with maximum energy about $250$ MeV are created by LWFA mechanism and it keeps a good shape (see Fig. \ref{fig2} (c)) and maximum energy more than $200$ MeV when it appears behind the foil. This also indicate that the electron beam almost unaffected by the reflected laser pulse and the foil. This resulting the pure electron beam and laser pulse interaction process, which has some potential applications in the strong laser and electron beam interaction studies. It is worth to noted that, during the reflected laser pulse interacting with the electron beam, the electrons are decelerated by the laser field and emit $\gamma$ ray photons via inverse Compton scattering, this resulting in a decrease in maximal energy of electrons some extent, as shown the inset in Fig. \ref{fig3} (a).

Figure \ref{fig3}(b) and (c) are plotted the electron beam energy and angular distributions at times t=225 fs and t=420 fs, respectively. The wakefield accelerated electron beam are mainly forward propagating after interaction with the reflected laser, and its maximal energy still about 120 MeV but with relatively bigger opening angle, as shown in Fig. \ref{fig3}(b). However, at time t=420 fs, the electron beam changes its direction and co-prorogating with the colliding laser pulse and directly accelerated by the laser pondremotive force to very high energies with smaller opening angle, as shown in Fig. \ref{fig3}(c). This is important for $\gamma$ ray emission which will discussed in next section.

\section{Mechanism of $\gamma$ ray emission}\label{Sec4}

Now, we consider the properties of the $\gamma$ ray emission process by the wakefield accelerated electron beam with the reflected and colliding laser pulses in more detail. High energy $\gamma$ photons can be emitted via nonlinear Compton scattering during high energy wakefield accelerated electrons interacting with a strong laser field. Quantum effects for electron dynamics is governed by the dimensionless and relativistic invariant parameter $\chi_e=|F_{\mu \nu}P^{\nu}|/{E_{s}m_{e}}$, where $F_{\mu \nu}=\partial_\mu A_\nu-\partial_\nu A_\mu$ is the electromagnetic field four-tensor, $P^{\nu}=\gamma m_e(c,\mathbf{v})$ is the electron four-momentum and $E_s=m_e^2/e$ is the critical Schwinger field, and the Plank units ($\hbar$ = c = 1) are used above.  Physically, $\chi_e$ is the ratio of the electric field in the electron rest frame to the critical Schwinger field. In the case of the a plane wave, which propagating in positive $x$ direction, It can be expressed as $\chi_e=(E/E_s)(\gamma - p_x/m_e)$, and the quantum effect is take in to account only if $\chi_e > 1$. The required laser intensity is very high and cannot be reached with realistic lasers. However, a relativistic electron beam counter-propagating with the laser field has $\chi_e ¡Ö 1$ due to the Lorentz boost, then $\gamma$ emission become important. For example, in the case of a relativistic plane electromagnetic wave counter-propagating with an electron beam, the invariant parameter is maximized, i.e., $\chi_e = 2\gamma(E/E_s)$, and the probability of the $\gamma$ photon emission is enhanced. In the contrary, for the co-propagating case it decrease as $\chi_e = (2\gamma)^{(-1)}(E/E_s)$, and the probability of the $\gamma$ photon emission is decreased dramatically\cite{Gu1,Gu2}. The details of $\gamma$ ray emission are discussed in the following subsections.

\subsection{$\gamma$ ray emission by wakefield electron interacting with a counter-propagating laser pulse}
In this subsection we begin our discussion from a QED-PIC simulations results of a $\gamma$ ray emission process, which includes both of the electron interacting with reflected laser from the foil and interacting another high intensity laser pulse behind the foil. In the simulation we have considered an underdense preplasma with the density of $n=0.05n_c$, and the circularly polarised laser pulses are used both for driving and colliding lasers but with different intensities.

\begin{figure}[tbp]\suppressfloats
\centerline{\includegraphics[width=10cm]{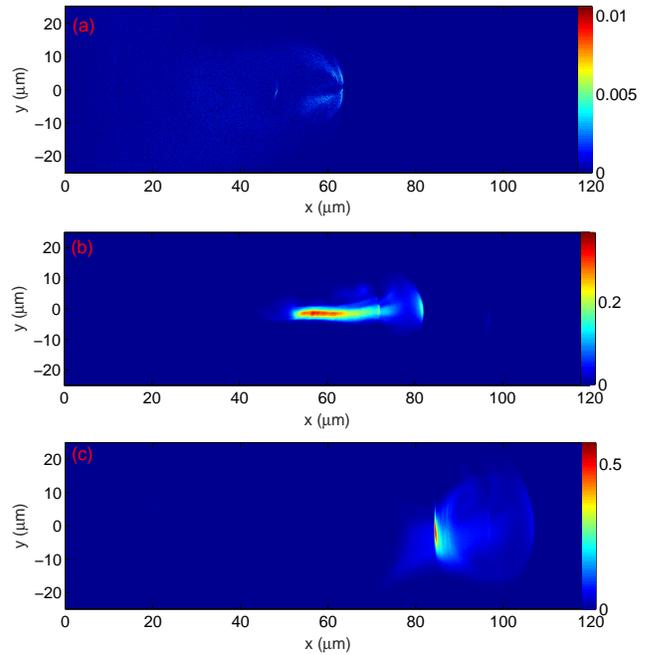}}
\caption{\label{fig4} Spatial distribution of the photon densities in the ($x, y$) plane at $t=225$ fs (a), at $t=337$ fs, and at $t=420$ fs (c).}
\end{figure}
\begin{figure}[tbp]\suppressfloats
\centerline{\includegraphics[width=8cm]{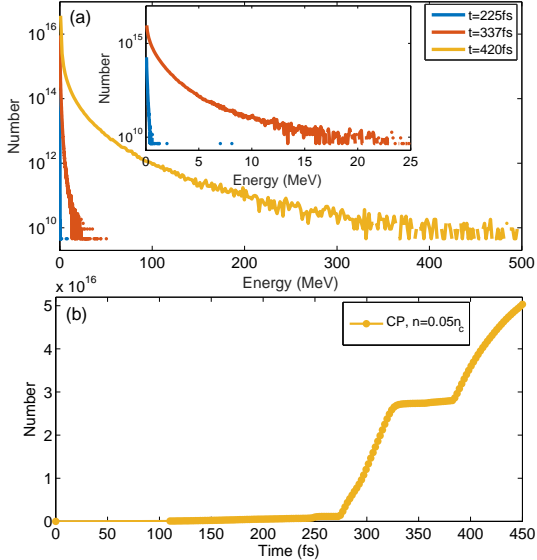}}
\caption{\label{fig5} (a) The energy spectrum of $\gamma$ ray photons at $t=225$ fs (blue curve), $t=337$ fs (red curve), and $t=420$ fs (yellow curve), respectively. (b) The evolution of the number of $\gamma$ ray photons with time. }
\end{figure}

Figure \ref{fig4} displays the spatial distributions of the $\gamma$ ray photons density in the ($x, y$) plane at different times. We can see from the Fig. \ref{fig4}(a) that there are few $\gamma$ ray photons produced during bubble formation. In this case, electrons in preplasma accelerated via direct laser acceleration mechanism, and the electrons transverse velocity is the order of the speed of light. Consequently, the electrons experience Betatron oscillation and emit some low energy $\gamma$ photons. Besides, it has large space distribution characteristics and the maximum energy is only about 1 MeV because of the lower intensity of the driving laser pulse which only about $a=20$, as shows in Fig. \ref{fig5} (a) and Fig. \ref{fig7} (a).

One can clearly see from Fig. \ref{fig4} (b) that a long $\gamma$ ray flash can be produced in the bubble by wakefield accelerated electrons interacting with reflected laser pulse, and its maximum energy reaches to above 25 MeV with photon numbers more than $10^{11}$ as shows in Fig. \ref{fig5}(a). In this stage, the reflected laser field from the foil interacting with the wakefield accelerated electron beam, i.g., the electron beam counter-propagating with the reflected laser field. In this scenario the $\chi_e$ is maximized, and the chance of the $\gamma$ photon generation and the photon energies are increased dramatically. Besides, because of the electron beam is long and the reflected laser cannot compress the electron beam in this stage, the electron beam still keep its long shape. Therefore, the reflected laser pulse acts like a natural undulator and a long $\gamma$ ray flash is generated. The length of the $\gamma$ ray beam also can be controlled via controlling the LWFA process with adjusting the plasma and laser parameters.

Figure \ref{fig4} (c) shows that a short $\gamma$ ray flash can be produced by the forward propagating electron beam interacting with a counter propagating high intensity laser pulse behind the foil. It shows that the $\gamma$ ray maximum energy is exceeds 400 MeV and number of the emitted photons are still above $10^{11}$ as shows in Fig. \ref{fig5} (a). In this stage, the laser field is very intense and all of the electrons are stopped and reflected by the strong laser field and compressed in a small space, and then accelerated to high energy again in the positive $x$ direction. Consequently, a very short $\gamma$ rays flash with high energies are produced in the interaction.

After the first stage is over, the wakefield accelerated electron beam still keep its energy and shape, going through the foil and interacting a more intense colliding laser pulse In the second stage. Figure \ref{fig5}(a) depicts the energy spectrum of the $\gamma$ ray photons in the first and second stages. We can clearly see that the photons maximal energy increases factor of 10 after the electron beam interacting with reflected laser pulse. Moreover, the maximum photon energy up to 500 MeV after the collision of the electron beam with a counter propagating laser pulse behind the foil. Figure \ref{fig5}(b) shows that the evolution of the number of $\gamma$ ray photons with time. At the beginning of the LWFA process, there are few low energy photons by electron Betatron oscillation in the bubble field as mentioned in the Fig. \ref{fig4}(a). When the electron beam interacting with the reflected laser pulse in the first stage, a copious of photons are emitted, the total yield of photons is reached about $3\times10^{16}$ and keeps it as a while. However, in the second stage, the total yield of photons reaches to $5\times10^{16}$ at time $t=420$ fs.
\begin{figure}[tbp]\suppressfloats
\centerline{\includegraphics[width=8cm]{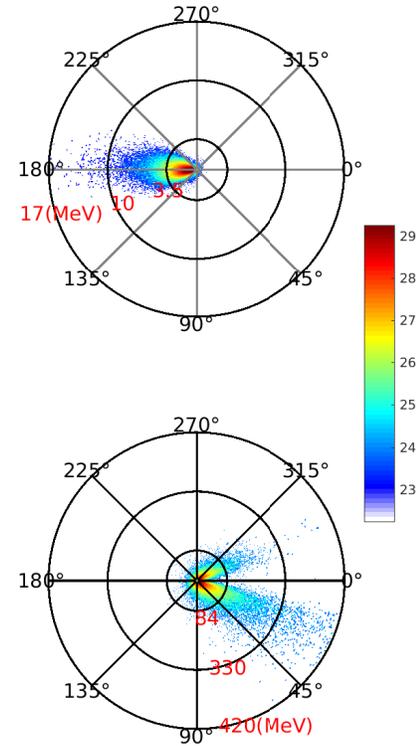}}
\caption{\label{fig6} Energy angular distributions of photons at t=337 fs (a) and t= 420 fs (b), respectively. The radial direction represents the energy, and the color bar shows the electron number in $\log$ scale.}
\end{figure}

Figure \ref{fig6} presents the photons energy angular distribution at $t=337$ fs and $t=420$ fs, respectively. It shows that the photons emitted in opposite directions in first and second stages, as shown in Fig. \ref{fig6} (a) and (b). The important point is that, in the second stage, the electron beam with energies about 200 MeV head on colliding with the high intensity counter-propagating laser pulse. The electron reflected immediately by the colliding laser, and accelerated to opposite direction by the strong ponderomotive force of the laser  (see Fig. \ref{fig2} and Fig. \ref{fig2}), then the counter-propagating situation is became a co-propagating immediately, results the minimizing the $\chi_e$. This also resulting the oppositeness of the angular energy distribution. Moreover, the electrons neither entering nor staying in the high field region in second stage, this also results decreasing the probability of the photon emission some extent.

\subsection{Effect of the driving laser pulse polarization on the $\gamma$ ray emission efficiency}

\begin{figure}[tbp]\suppressfloats
\centerline{\includegraphics[width=8cm]{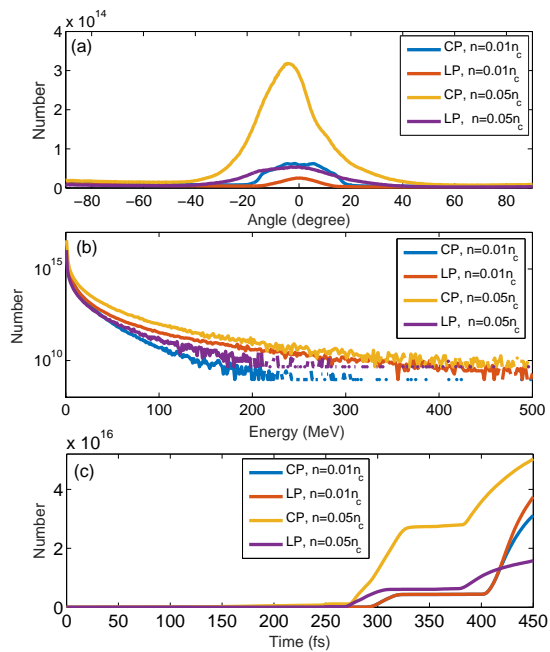}}
\caption{\label{fig7} (a) The angular distribution of the $\gamma$ ray photons at $t=420$ fs for different driving laser polarization and different preplasma densities of $n=0.01n_c$ and $n=0.05n_c$. (b) The corresponding energy spectrum of $\gamma$ photons at $t=420$ fs. (c) Evolution of the $\gamma$ ray photons number with time for different driving laser polarization and different preplasma densities of $n=0.01n_c$ and $n=0.05n_c$}
\end{figure}
In this subsection, we investigate the effects of the driving laser pulse polarization on the $\gamma$ ray emission efficiency. In our simulations we consider linearly and circularly polarized laser pulses and two kinds of preplasma with density of $n=0.01n_c$ and $n=0.05n_c$ while the foil density is fixed to $n=30n_c$. The more different preplasma density configurations are investigated in the next subsection in detail. Figure \ref{fig7}(a) depicts the angular distribution of the $\gamma$ ray photons at $t=420$ fs. One can clearly see that more $\gamma$ ray photons could be generated in the second stage for the circularly polarized laser case when the preplasma density are chosen as $n=0.05n_c$. In this case, the HWFM of divergence angle is about $20^\circ$ and corresponding total photon number about $10^{14}$. Figure \ref{fig7}(b) and (c) present the energy spectrum and the evolution of the $\gamma$ photon numbers with time $t$, respectively. The results also indicate that for slightly higher preplasma density£¬the circular polarized laser pulse has the great advantages to generate more $\gamma$ ray photons with smaller divergence angle. It is because, for the higher preplasma density, greater bubble field can be build and more electrons could be trapped and accelerated. Besides, the circular polarized laser pulse could effectively suppress the electrons transverse oscillation and heating, and then more electrons concentrate at the bubble center. Consequently, the more energetic electrons can escape from the deceleration field in the bubble front and sheath electric field rear the foil more easily. Then, more electrons collide with the counter propagating laser pulse in the second stage and emit more $\gamma$ ray photons.

\subsection{Effect of the preplasma density on the $\gamma$ ray emission efficiency}

\begin{figure*}[tbp]\suppressfloats
\centerline{\includegraphics[width=11cm]{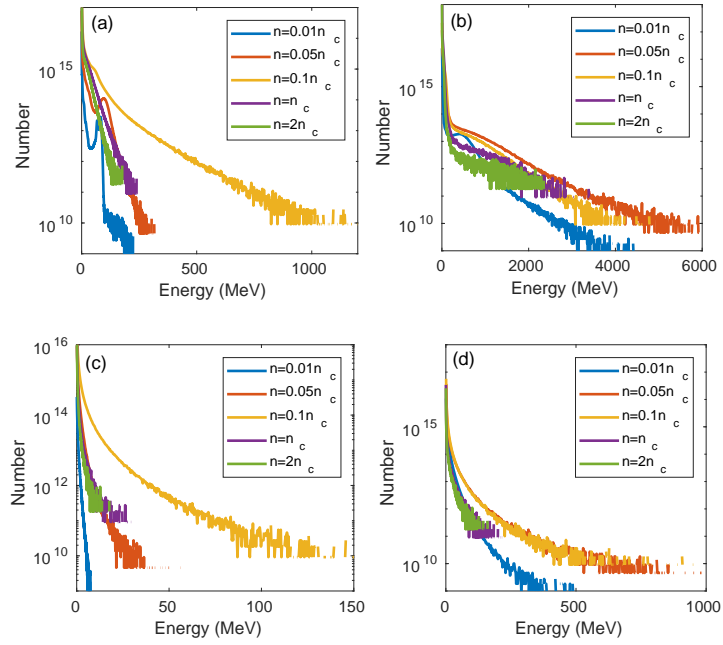}}
\caption{\label{fig8} The energy spectra of the electrons (top row) and photons (bottom row) for different preplasma densities in the different stages. (a) and (b) The electrons energy spectra at t=337 fs and t=420 fs, respectively. (c) and (d) The $\gamma$ ray photons energy spectra at t=337 fs and t=420 fs, respectively.}
\end{figure*}
We can conclude from previous subsection that the proper choose of the preplasma density is very important for efficient $\gamma$ photon emission. Therefore, we investigate the effects of the preplasma density on the efficiency of $\gamma$ ray emission. In our simulations we consider circularly polarized laser pulses and five kinds of preplasma with densities of $n=0.01n_c$, $n=0.05n_c$, $n=0.1n_c$, $n=n_c$ and $n=2n_c$ while the foil density is fixed to $n=30n_c$. Figure \ref{fig8} depicts the energy spectra of the electrons (upper row) and $\gamma$ photons (bottom row) at $t=337$ fs (left column) and $t=420$ fs (right column), respectively. One can clearly see that, in the first stage, the wakefield accelerated electrons maximal energy increasing with the preplasma density and reach the optimal when $n=0.1n_c$, and then decreasing with the preplasma density as shown in the Fig. \ref{fig8} (a). Accordingly, the $\gamma$ ray photons energy reaches about to 150MeV in the first stage with preplasma density $n=0.1n_c$, as shown in Fig. \ref{fig8} (c). However, in the second stage, there have no big difference both of the electrons and photons energy spectra in the $n=0.05n_c$ and $n=0.1n_c$ cases as shown in the Fig. \ref{fig8} (b) and (d). The conclusions of these results that, for the higher preplasma density, greater bubble field can be build and more electrons trapped and accelerated by this field in the first stage. Consequently, more electrons collide with reflected laser pulse and emit more $\gamma$ ray photons compared with lower density preplasma case. However, for the higher preplasma density, the laser intensity is not high enough to build an efficient wakefield, this directly decreases the efficiencies of the electron acceleration and the $\gamma$ photon generation. Thus, the preplasma with the density of $n=0.1n_c$ has the great advantages in the first stage in our percent scheme, where the $a_0=20$ for driving laser. We all know that, very low density preplasma not recommended to wakefield electron acceleration, because of the weaker wakefield and fewer electrons are trapped and accelerated. This also reduced the efficiency of the $\gamma$ photons emission. We could be get more energetic electrons and photons via increasing the preplasma density to several times of critical density but the costs of the higher driving laser intensity. However, in the second stage of the acceleration, the more energetic electrons reflected harder and backward accelerated by the colliding laser pulse. Thus the electron maximum energy for the $n=0.1n_c$ case is smaller than $n=0.05n_c$ case. Accordingly, more energetic $\gamma$ photons are emitted in case of the $n=0.05n_c$ in the second stage.

\subsection{$\gamma$ ray emission with and without the foil}
\begin{figure*}[tbp]\suppressfloats
\centerline{\includegraphics[width=10cm]{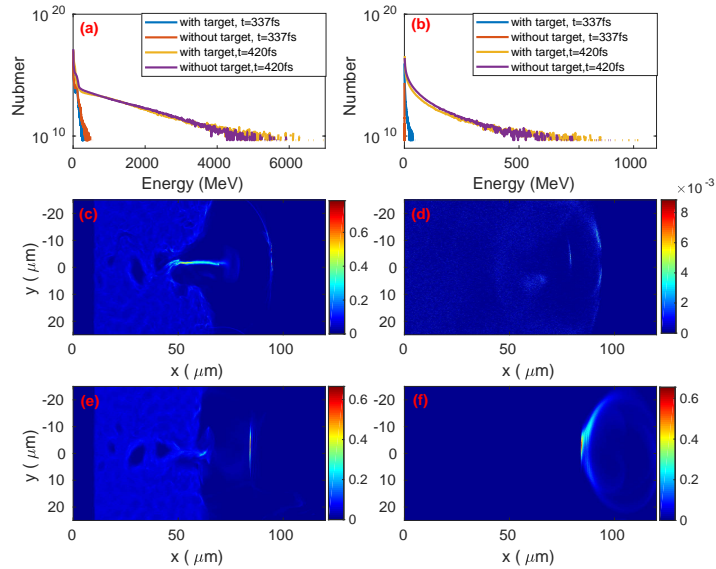}}
\caption{\label{fig9} The energy spectra and spatial distributions of the electrons and $\gamma$ ray photons with and without target cases at t=337 fs and t=420fs, respectively. (a) The electrons energy spectra at t=337 fs and t=420 fs, respectively. (b) The $\gamma$ ray photons energy spectra at t=337 fs and t=420 fs, respectively. The spatial distributions of the $\gamma$ ray photons with the foil at t=337 fs (c) and t=420 fs (e). The spatial distributions of the $\gamma$ ray photons without the foil at t=337 fs (d) and t=420 fs (f).}
\end{figure*}

\begin{figure*}[tbp]\suppressfloats
\centerline{\includegraphics[width=10cm]{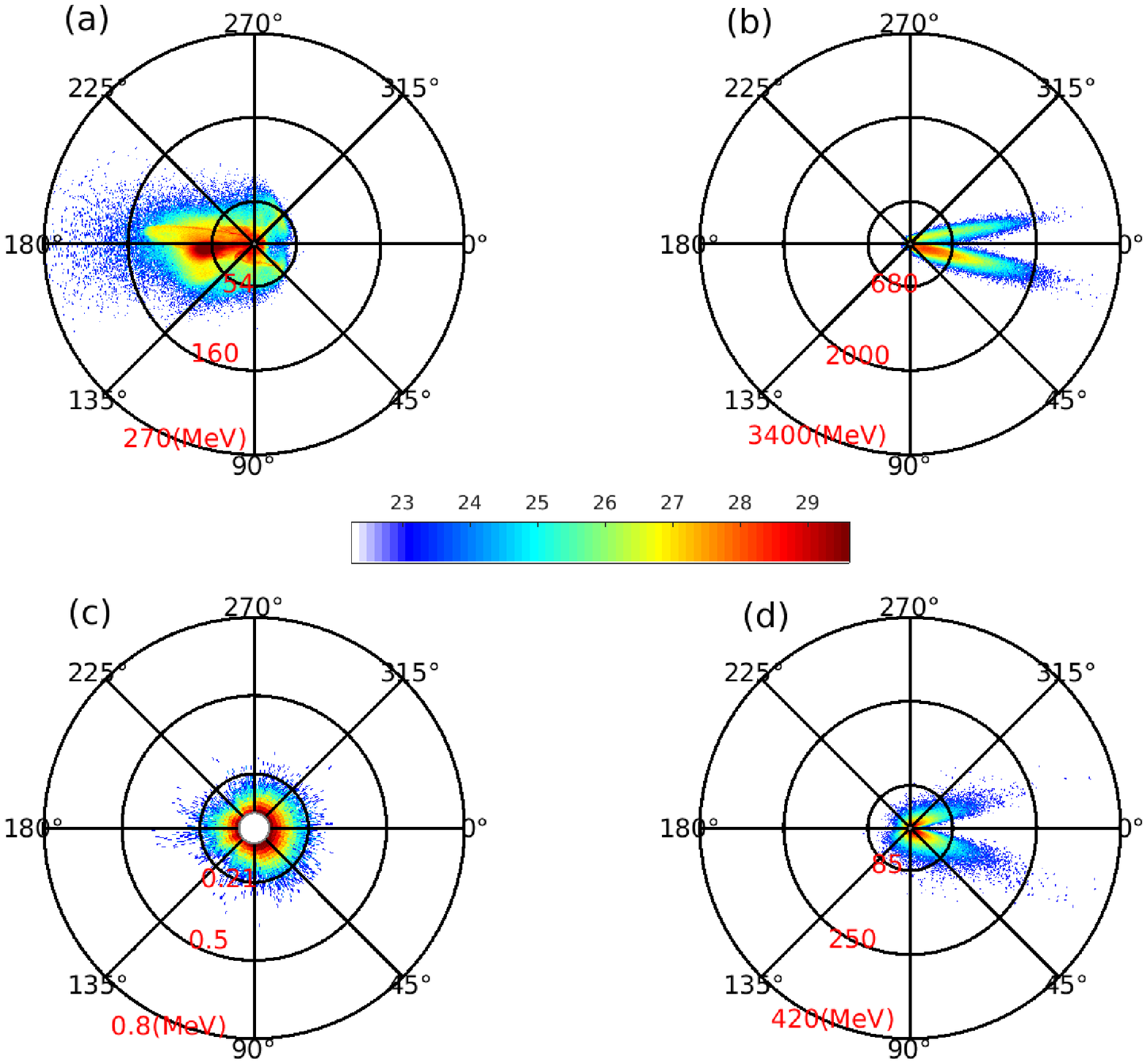}}
\caption{\label{fig10} Energy angular distributions of the electrons (top row) and $\gamma$ photons (bottom row) for the without foil case at t=337 fs [(a) and (c)], and t= 420 fs [(b) and (e)], respectively. The radial direction represents the energy and the color bar shows the electron number in $\log$ scale. }
\end{figure*}
Finally, we investigate the effects of the foil on the efficiency of $\gamma$ ray emission in the second stage. We carried out two sets of simulations with and without the solid foil end of the preplasma, and the simulation results are show in Fig. \ref{fig9} and Fig. \ref{fig10}. In our simulations we consider circularly polarized laser pulses and preplasma with density of $n=0.05n_c$ according to the results of previous subsections. Figure \ref{fig9}(a) and (b) plots the energy spectra of the electrons and $\gamma$ ray photons, respectively. The electrons spectra are the similar for with and without foil cases in the first and second stage. However, the $\gamma$ photons energy is about zero for without target in first stage, and there have a similar photon energy spectra both of with and without foil cases in the second stage, as shown in Fig. \ref{fig9} (b). The spatial distribution of the $\gamma$ photons with (left) and without (right) the foil are plotted in the Fig. \ref{fig9} (c) and (d). There are relatively long $\gamma$ ray flash with target case, because there has a reflected laser pulse interacting with the electrons. In the contrary, there have few low energy photons in the first stage for the no foil case. The photons densityare also similar in the both with and without target cases at t=420 fs, but there is small difference its spatial distributions as shown in Fig. \ref{fig9} (e) and (f). In the Fig. \ref{fig10} we plot the angular energy distributions of the electrons (top row) and photons (bottom row) for the without foil case at t=337 fs (left) and t=420 fs (right), respectively. As can be seen in the Fig. \ref{fig10} (a) and (b) that, the electrons forward and backward propagating at the different stages. The $\gamma$ photons energy at t=337 fs has the very broad angular distribution and its energies are very low as shown as Fig. \ref{fig10} (c). This $\gamma$ photons mainly emitted via Betatron oscillation of the direct laser accelerated electrons, rather than by wakefield accelerated electrons interacting with the reflected laser field. However, in the second stage, a backward propagating $\gamma$ ray photons are emitted, and the angular energy distribution characters are similar to the with foil case, as shown in Fig. \ref{fig10} (d) and Fig. \ref{fig6} (b).

\section{Discussion and Conclusion}\label{Sec5}
In this paper, we have studied an efficient method to generation of the $\gamma$ ray photons by the interaction a wakefield accelerated electron beam with a counter propagating laser pulses. In our simulations, we used the compound plasma target with an underdense preplasma and a solid foil attached to the preplasma. We have considered the different preplasma densities while the foil density is unchanged. The circularly and linearly polarized driving laser pulses are used in our studies to induce the wakefield. Our numerical simulations are performed by using the QED-PIC simulation code EPOCH. The simulation results show that bubble structure can be developed by the interaction of an intense short circularly/linearly laser pulses with the preplasma. The electrons trapped and accelerated in the bubble to high energies. As the laser pulse reflected by the solid foil, the accelerated electrons can passing behind the foil and appear at the back of the foil as an energetic electron beam with little energy loss and divergence. Consequently, twin $\gamma$ ray beams with the time delay are generated front and back of the solid foil. Besides, we compare the simulation results different preplasma densities, driving laser polarization and with /without solid foil cases. It should be mentioned that in the first stage, the efficiency of the $\gamma$ ray photon emission can be enhanced with the foil, and it dramatically decreased without the foil. In the contrary, in the second stage, the photon emission efficiencies are similar in the cases of with and without foil. As the same time, our approach has the advantages of that£¬the delay and energy could be fine-tuned by adjusting the plasma and laser parameters accordingly, which might be useful for future QED investigations.

In conclusion, the 2D QED-PIC simulation results are indicated that, two groups of $\gamma$ ray beams with tunable energy and delay are produced in this method. During the interaction between the wakefield accelerated electrons and reflected laser pulse from the foil, a long $\gamma$ ray is produced. As the time goes on, a short $\gamma$ ray with the divergence angle about $20^\circ$ and photon number about $10^{14}$ are produced by the interaction of an counter propagating high-intensity laser pulse with the forward traveling electron beam behind the foil. It is also found that, the efficiency of the $\gamma$ ray emission is closely related to the driving laser polarization, preplasma density and the solid foil as well. We found that slightly higher preplasma density and circular polarized laser pulse has the great advantages to generate more $\gamma$ ray photons with smaller divergence angle. Also, the efficiency of the $\gamma$ ray emission in the first stage greatly related to foil, while in the second stage it less related to the foil.  Namely, $\gamma$ ray maximum energy, number and energy spread are greatly enhanced when the preplasma parameters and laser polarization are chosen properly.

\begin{acknowledgments}
he authors would like to thank Dr. N. Abdukerim for her helpful discussions. This work was financially supported by the National Natural Science Foundation of China (NSFC) (GrantNos. 11664039,11575150, 11964038 and 11875007). The authors are particularly grateful to CFSA at the University of Warwick for allowing us to use the EPOCH.
\end{acknowledgments}


\begin{thebibliography}{99}\suppressfloats

\bibitem{Tajima}
Tajima T \textit{et al} 1979 \textit{Phys. Rev. Lett.} \textbf{43} 267

\bibitem{Malka}
Malka V \textit{et al} 2002 \textit{Science} \textbf{298} 5598

\bibitem{Wang}
Wang X M \textit{et al} 2013 \textit{Nat. Commun.} \textbf{4} 1988

\bibitem{Leemans}
Leemans W P \textit{et al} 2014 \textit{Phys. Rev. Lett.} \textbf{113} 245002

\bibitem{Gonsalves}
Gonsalves A J \textit{et al} 2019 \textit{Phys. Rev. Lett.} \textbf{122} 084801

\bibitem{Powers}
Powers N D \textit{et al} 2013 \textit{Nat. Photonics} \textbf{8}, 28

\bibitem{Lemos}
Lemos N \textit{et al} 2019 \textit{Phys. Plasmas} \textbf{26} 083110

\bibitem{Chenmin}
Luo J \textit{et al} 2016 \textit{Scientific Rep.} \textbf{6} 29101

\bibitem{Albert}
Albert A \textit{et al} 2016 \textit{Plasma Phys. Control. Fusion} \textbf{58} 103001

\bibitem{Petawatt}
Danson C \textit{et al} 2015 \textit{High Power Laser Sci. Eng.} \textbf{3} e3

\bibitem{ELI}
ELI Beamlines, www.eli-beams.eu.

\bibitem{ican}
Mourou G \textit{et al} 2013 \textit{Nat. Photonics} \textbf{7} 258

\bibitem{vulcan}
The vulcan 10 petawatt project,
http://www.clf.stfc.ac.uk/CLF/Facilities/
Vulcan/The+Vulcan+10+Petawatt+Project/14684.aspx

\bibitem{Yan}
Yan W C \textit{et al} 2017 \textit{Nat. Photonics} \textbf{11} 514

\bibitem{Benedetti}
Benedetti A \textit{et al} 2018 \textit{Nat. Photonics} \textbf{12} 319

\bibitem{Vranic}
Vranic M \textit{et al} 2018 \textit{Scientific Rep.} \textbf{8} 4702

\bibitem{BakeQED}
Bake M A \textit{et al} 2018 \textit{Front. Phys.} \textbf{13} 135202

\bibitem{Chen}
Chen S \textit{et al} 2013 \textit{Phys. Rev. Lett.} \textbf{110} 155003

\bibitem{Sarri}
Sarri G \textit{et al} 2014 \textit{Phys. Rev. Lett.} \textbf{113} 224801

\bibitem{Lobet}
Lobet M \textit{et al} 2014 \textit{Phys. Rev. ST Accel. Beams} \textbf{20} 043401

\bibitem{Chang}
Chang H X \textit{et al} 2017 \textit{Scientific Rep.} \textbf{7} 45031

\bibitem{Yu}
Yu J Q \textit{et al} 2019 \textit{Phys. Rev. Lett.} \textbf{122} 014802

\bibitem{Xie}
Hou Y J \textit{et al} 2019 \textit{Plasma Sci. Technol.} \textbf{21} 085201

\bibitem{Cipiccia1}
Cipiccia S \textit{et al} 2012 \textit{J. Appl. Phys.} \textbf{111} 063302

\bibitem{Cipiccia2}
Cipiccia S \textit{et al} 2011 \textit{Nat. Phys.} \textbf{7} 867

\bibitem{Phuoc}
Phuoc K \textit{et al} 2012  \textit{Nat. Photonics} \textbf{6} 308

\bibitem{Wenz}
Wenz J \textit{et al} 2019  \textit{Nat. Photonics} \textbf{13} 263

\bibitem{Liupop}
Liu J B \textit{et al} 2019 \textit{Phys. Plasmas} \textbf{26} 033109

\bibitem{Gu1}
Gu Y J \textit{et al} 2018 \textit{Opt. Express} \textbf{26} 19932

\bibitem{Gu2}
Gu Y J \textit{et al} 2018 \textit{Communications Phys.} \textbf{1} 93

\bibitem{Thomas}
Thomas A G R \textit{et al} 2012 \textit{Phys. Rev. X} \textbf{2} 041004

\bibitem{Liu}
Liu J X \textit{et al} 2016 \textit{Plasma Phys. Control. Fusion} \textbf{58} 125007

\bibitem{JMCole}
Cole J M \textit{et al} 2018 \textit{Phys. Rev. X} \textbf{8} 011020

\bibitem{Poder}
Poder K \textit{et al} 2018 \textit{Phys. Rev. X} \textbf{8} 031004

\bibitem{EPOCH}
Arber T D \textit{et al} 2015 \textit{Plasma Phys. Control. Fusion} \textbf{57} 113001
\end{thebibliography}
\end{document}